\documentclass[aps,prl,showpacs,twocolumn]{revtex4}
\usepackage{graphicx}
\newcommand{\U}{{\cal U}}
\begin{document}
\draft
\title{Unparticles in Gluodynamics}
\author{A.T. Alan}
\email[ alan\_a@ibu.edu.tr]{}
 \affiliation{Department of
Physics, Abant Izzet Baysal University, 14280 Bolu, Turkey}
\begin{abstract}
The virtual effects of unparticle physics on the two-gluon jets
production are explored for CERN LHC. It is found that depending on
the scale dimension, unparticles can give rise to substantial
enhancements in the transverse momentum and pseudo-rapidity
distributions of hadronic cross sections, up to a factor of
four-orders of magnitude.
\end{abstract}
\maketitle
{\it Introduction}. Recently, Georgi \cite{Georgi:2007ek} has
introduced the concept of unparticle physics as a scale invariant
field theory, which proved to be very interesting and attractive as
displayed by the number of publications on the subject \cite{2}.
Fields of unparticle physics do not manifest themselves as particles
unlike the fields of conventional field theories. The scheme is as
follows. The fields of a scale invariant sector with a nontrivial
infrared fixed point -Banks and Zaks ($\mathcal{BZ}$) fields-
\cite{Banks:1981nn} and the fields of Standard Model (SM) can
coexist and interact via the exchange of particles of large mass
scale $M_\U$ at very high energies. Below $M_\U$, non-renormalizable
operators are induced giving the interactions
\[\frac{1}{M_\U^k}\mathcal{O}_{SM}\mathcal{O}_{\mathcal{BZ}}\]
in the generic form. Where $\mathcal{O}_{SM}$ and $\mathcal{O}_{BZ}$
are operators built out of $SM$ and $\mathcal{BZ}$ fields with scale
dimensions $d_{SM}$ and $d_{BZ}$, respectively. Renormalizable
couplings of $\mathcal{BZ}$ fields then cause dimensional
transmutation as scale invariance emerges at an energy scale
$\Lambda_\U$. Below $\Lambda_\U$, $\mathcal{BZ}$ operators match
onto unparticle operators giving the non-renormalizable interactions
\[C_\U\frac{\Lambda_\U^{d_{\mathcal{BZ}}-d_\U}}{M_\U^k}\mathcal{O}_{SM}\mathcal{O}_{\U}\]
where the non-integral number $d_\U$ is the scaling dimension of the
unparticle operator $\mathcal{O}_\U$ constructed out of the new
fields (unparticles) and $C_\U$ is a coefficent function.
Following this suggestion many studies have been performed to
explore the effects of unparticles on low energy dynamics \cite{UP}.
Our aim in this letter is to explore the effects of virtual
unparticles on the two-gluon jets production which will be one of
the most observed processes at the forthcoming LHC experiments.

At hadron colliders jet production predominates all the other
processes. Its understanding in detail is crucial both for precision
tests of perturbative Quantum Chromodynamics (pQCD) and the search
for new physics. Data from the experiments at the Tevatron are in
good agreement with the next-to-leading order (NLO) pQCD predictions
\cite{Acosta:2004js,Campbell:2006wx}. Specifically the LHC, with 14
TeV center of mass energy and a high luminosity of $10^5$ $pb^{-1}$,
will provide very large accessible kinematic range, which in turn
will present unprecedented discovery potential. Cross section
measurements at the LHC will cover jets with transverse momenta of
order 4 TeV to probe the shortest distances ever reached.

{\it Two-gluon jets at hadronic collisions}. In this letter, we
exploit the implications of unparticle physics on two-gluon jets
production in $pp$ collisions via the partonic $gg\rightarrow gg$
scattering at LO, in the framework of gluodynamics which describes
gluons and their selfinteractions only. Gluon jets show differences
compared to quark jets in their widths, hadron multiplicities and
ratios of multiplicities stemming from different color charges of
quarks and gluons. Great progress has been made on these differences
by both theorists and experimentalists \cite{Jets}. As we do not
expect a leading contribution for this production, the analysis of
the $q\bar q$ initial state will not be included in present work.
Gluon fusion as an initial state is largely dominant compared to
those of quark-antiquark annihilation and quark-gluon scattering in
most of the production channels at the LHC energies, indeed.

\begin{figure}[httb!]
 \includegraphics[width=7cm]{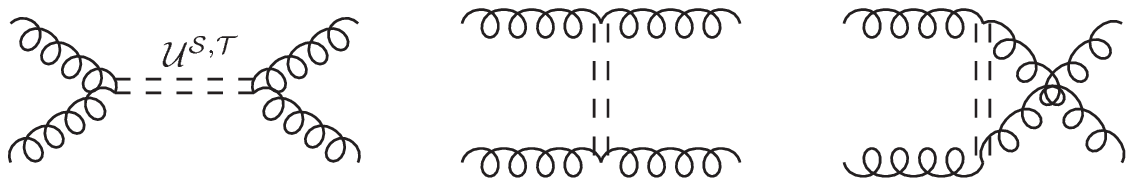}\\\includegraphics[width=8cm]{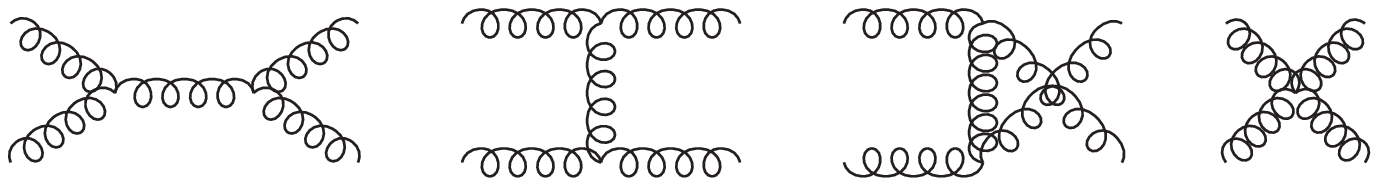}
  \caption{The leading order Feynman diagram for $gg\rightarrow gg$ scattering with the contribution of unparticles.}\label{fig1}
\end{figure}

The relevant tree level Feynman graphs are shown in Fig.~\ref{fig1}
in which the upper three graphs correspond to the contribution of
virtual scalar and tensor unparticles, as vector unparticles do not
couple to the gluons. The lower four ones are the usual tree level
SM diagrams.
To derive the corresponding differential cross sections we work in
the axial gauge by taking the gluon propagator as follows:
\begin{eqnarray}
\Delta_{\mu\nu}=\frac{-i}{q^2}(g_{\mu\nu}-\frac{q^{\mu}q^{\nu}}{q^2}).
\end{eqnarray}
In summing over the physical, transverse gluon polarizations in the
calculations of  amplitudes squared, we use
\begin{eqnarray*}
\sum_{\lambda}\epsilon_{\mu}(\lambda,p)\epsilon_{\nu}^{\star}(\lambda,p)&=&(-g_{\mu\nu}-\frac{n^{\mu}p^{\nu}+n^{\nu}p^{\mu}}{n\cdot
p})
\end{eqnarray*}
where the fourvector $n$ is defined as $n=(p_0,-\vec{p})$.
Propagators for the scalar and the tensor unparticles
\cite{Georgi:2007ek,Cheung:2007ap}, derived from a scale invariance
principle, are
\begin{eqnarray}\label{prou}
\Delta^S(q^2_{\U})&=&\frac{iA_{d_\U}}{2\sin(\pi
d_{\U})}(-q^2_{\U})^{d_{\U}-2}\nonumber\\
\Delta_{\mu\nu,\rho\sigma}^T(q^2_{\U})&=&\frac{iA_{d_\U}}{2\sin(\pi
d_{\U})}(-q^2_{\U})^{d_{\U}-2}T_{\mu\nu,\rho\sigma}(q)
\end{eqnarray}
respectively, where
\begin{equation}
   A_{d_\U}={16\pi^2\sqrt{\pi}\over (2\pi)^{2{d_\U}}}
       { \Gamma({d_\U}+{1\over
       2})\over\Gamma({d_\U}-1)\Gamma(2\,{d_\U})}  \;
\end{equation}
is the normalization factor of $d_{\U}$-body phase space of massless
particles and the tensor $T^{\mu\nu,\rho\sigma}$ in Eq.~(\ref{prou})
is given by
\begin{eqnarray}\label{eq4}
T_{\mu\nu,\rho\sigma}(q) &=& \frac{1}{2} \, \{
   \pi_{\mu\rho}(q)\  \pi_{\nu\sigma}(q)
        + \pi_{\mu\sigma}(q)\  \pi_{\nu\rho}(q)\nonumber\\ &-& \frac{2}{3}\
          \pi_{\mu\nu}(q)\  \pi_{\rho\sigma}(q)  \} \;
          \end{eqnarray}
with
\begin{eqnarray}
 \pi_{\mu\nu}(q) &=& - g_{\mu \nu} + \frac{q_\mu q_\nu }{ q^2} \;
\end{eqnarray}
for transverse and traceless tensor unparticle operator. In
conformal field theories (CFT) this tensor is given by
\cite{Grinstein:2008qk}
\begin{eqnarray}\label{eq6}
T_{\mu\nu,\rho\sigma}&=&\frac{1}{2}
\bigg[\left(g_{\mu\rho}g_{\nu\sigma}+\mu\leftrightarrow\nu\right)
+\frac{\left[4-d(d+1)\right]}{2d(d-1)}g_{\mu\nu}g_{\rho\sigma} \nonumber\\
\nonumber && -2\frac{(d-2)}{d}\left(g_{\mu\rho}\frac{k_\nu
k_\sigma}{k^2} +g_{\mu\sigma}\frac{k_\nu
k_\rho}{k^2}+\mu\leftrightarrow\nu\right)
\nonumber \\
&& +4\frac{(d-2)}{d (d-1)}\left(g_{\mu\nu}\frac{k_\rho
k_\sigma}{k^2} +g_{\rho\sigma}\frac{k_\mu k_\nu}{k^2}\right)\nonumber\\
&&+8\frac{(d-2)(d-3)}{d(d-1)} \frac{k_\mu k_\nu  k_\rho
k_\sigma}{(k^2)^2}\bigg]~ \;.
\end{eqnarray}
The structures of vertices \cite{Cheung:2007zz,Chen:2007qr} for the
effective interactions of scalar and tensor unparticles with gluons
are SM gauge invariant and are given, respectively, by
\begin{eqnarray}
\lambda_0 \frac{1}{\Lambda_\U^{d_\U} } G_{\alpha\beta}
G^{\alpha\beta} O_\U \; ~~\mathrm{and}~~ \; \lambda_2
\frac{1}{\Lambda_\U^{d_\U} } G_{\mu\alpha} G_{\nu}^{\;\alpha}
O_\U^{\mu\nu}
\end{eqnarray}
where $G^{\alpha\beta}$ denotes the gluon field strength and
$\lambda_0, \lambda_2$ are dimensionless coupling constants for the
scalar and the tensor unparticles, respectively.

After helicity and color averaging the partonic differential cross
section in the case of the scalar unparticle exchange is obtained as
\begin{widetext}
\begin{eqnarray}\label{csus}
\frac{d\hat\sigma^S}{d\hat t}&=&\frac{1}{256\pi\hat
s^2}\Bigg\{A_S^2\Big[\frac{\hat s^4}{(\hat
s)^{4-2d_{\U}}}+\frac{\hat t^4(\hat t^2+\hat u^2)^2}{\hat s^4(-\hat
t)^{4-2d_{\U}}}+\frac{\hat u^4(\hat t^2+\hat u^2)^2}{\hat s^4(-\hat
u)^{4-2d_{\U}}}\Big] +A_S^2\bigg\{\cos(\pi d_\U)\Big[\frac{\hat
t^2(\hat t^2+\hat u^2)}{8(\hat s)^{2-d_\U}(-\hat
t)^{2-d_\U}}\nonumber\\&+&\frac{\hat u^2(\hat t^2+\hat u^2)}{8(\hat
s)^{2-d_\U}(-\hat u)^{2-d_\U}}\Big]+\frac{\hat t^2\hat u^2(\hat
t^4+6\hat u^2\hat t^2+\hat u^4)}{8\hat s^4(-\hat t)^{2-d_\U}(-\hat
u)^{2-d_\U}}\bigg\}+12\pi A_S\alpha_s\bigg\{\frac{\cos(\pi
d_\U)}{(\hat s)^{2-d_\U}}\Big[\frac{1}{2\hat t}(\hat s^3-7\hat
s^2\hat u-8\hat s\hat u^2-2\hat u^3)\nonumber\\ &+&\frac{1}{2\hat
u}(\hat s^3-7\hat s^2\hat t-8\hat s\hat t^2-2\hat t^3)+\hat
s^2+2\hat t\hat u\Big]+\frac{1}{(-\hat t)^{2-d_\U}}\Big[\frac{\hat
t^2(\hat t^3-\hat u\hat t^2+\hat u^2\hat t-\hat u^3)}{2\hat
s^3}\nonumber\\&-&\frac{\hat t^2(2\hat s^5+9\hat u\hat s^4+20\hat
u^2\hat s^3+20\hat u^3\hat s^2+16\hat u^4\hat s+8\hat u^5)}{2\hat
u\hat s^4}+\frac{\hat t^2(\hat s+2\hat t)^2(\hat s^2+\hat t\hat
s+\hat t^2)}{\hat s^4}\Big]\nonumber\\&+&\frac{1}{(-\hat
u)^{2-d_\U}}\Big[\frac{\hat u^2(\hat u^3+\hat u\hat t^2-\hat u^2\hat
t-\hat t^3)}{2\hat s^3}-\frac{\hat u^2(2\hat s^5+9\hat t\hat
s^4+20\hat t^2\hat s^3+20\hat t^3\hat s^2+16\hat t^4\hat s+8\hat
t^5)}{2\hat t\hat s^4}\nonumber\\&+&\frac{\hat u^2(\hat s+2\hat
u)^2(\hat s^2+\hat u\hat s+\hat u^2)}{\hat
s^4}\Big]\bigg\}\Bigg\}+\frac{d\hat\sigma_0}{d\hat t}
\end{eqnarray}
\end{widetext}
where
\begin{eqnarray}
A_S=\frac{8\lambda_0^2 A_{d_\U}}{\sin(\pi d_{\U})\Lambda^{2d_\U}}
\end{eqnarray}
 and
\begin{eqnarray}
\frac{d\hat\sigma_0}{d\hat t}=\frac{9\pi\alpha_s^2}{2\hat
s^2}\frac{(\hat s^2+\hat t\hat s+\hat t^2)^3}{\hat s^2\hat t^2\hat
u^2}
\end{eqnarray}
is the differential cross section for the SM prediction at LO.
$d\hat\sigma/d\hat t$ in Eq.~(\ref{csus}) consists of terms that
come from unparticle contributions ($A_S^2$ terms), interference of
gluons and unparticles ($A_S$ terms), as well as the SM term
$d\hat\sigma_0/d\hat t$. The corresponding differential cross
section for the tensor unparticle exchange with traceless tensor
operator in Eq.~(\ref{eq4}) is given by
\begin{widetext}
\begin{eqnarray}
\frac{d\hat\sigma^T}{d\hat t}&=&\frac{1}{32\pi\hat
s^2}\Bigg\{A_T^2\Big[\frac{\hat t^4+\hat u^4}{(\hat
s)^{4-2d_{\U}}}+\frac{\hat s^4+\hat u^4}{(-\hat
t)^{4-2d_{\U}}}+\frac{\hat s^4+\hat t^4}{(-\hat
u)^{4-2d_{\U}}}\Big]+\frac{A_T^2}{4}\bigg\{\cos(\pi
d_{\U})\Big[\frac{\hat u^4}{(\hat s)^{2-d_\U}(-\hat
t)^{2-d_\U}}+\frac{\hat t^4}{(\hat s)^{2-d_\U}(-\hat
u)^{2-d_\U}}\Big]\nonumber\\&+&\frac{\hat s^4}{(-\hat
t)^{2-d_\U}(-\hat u)^{2-d_\U}}\bigg\}+3\pi
A_T\alpha_s\bigg\{\frac{-\cos(\pi d_\U)}{(\hat
s)^{2-d_\U}}\Big[\frac{\hat t^5+2\hat u\hat t^4+\hat u^4\hat t+2\hat
u^5}{\hat t\hat s^2}+\frac{\hat u^5+2\hat t\hat u^4+\hat t^4\hat
u+2\hat t^5}{\hat u\hat s^2}-\frac{2(\hat t^4+\hat u^4)}{\hat
s^2}\Big]\nonumber\\&+&\frac{1}{(-\hat t)^{2-d_\U}}\Big[\hat s(\hat
t-\hat u)+\frac{2\hat t^5+13\hat u\hat t^4+28\hat u^2\hat t^3+25\hat
u^3\hat t^2+10\hat u^4\hat t+\hat u^5}{\hat u\hat s^2}+\frac{\hat
u^4+\hat s^2(2\hat u^2-\hat s^2-\hat t^2)}{\hat
s^2}\Big]\nonumber\\&+&\frac{1}{(-\hat u)^{2-d_\U}}\Big[\hat s(\hat
u-\hat t)+\frac{2\hat u^5+13\hat t\hat u^4+28\hat t^2\hat u^3+25\hat
t^3\hat u^2+10\hat t^4\hat u+\hat t^5}{\hat t\hat s^2}+\frac{\hat
t^4+\hat s^2(2\hat t^2-\hat s^2-\hat u^2)}{\hat
s^2}\Big]\bigg\}\Bigg\}+\frac{d\hat\sigma_0}{d\hat t}
\end{eqnarray}
\end{widetext}

\begin{figure}[htbp!]
 \includegraphics[width=9cm]{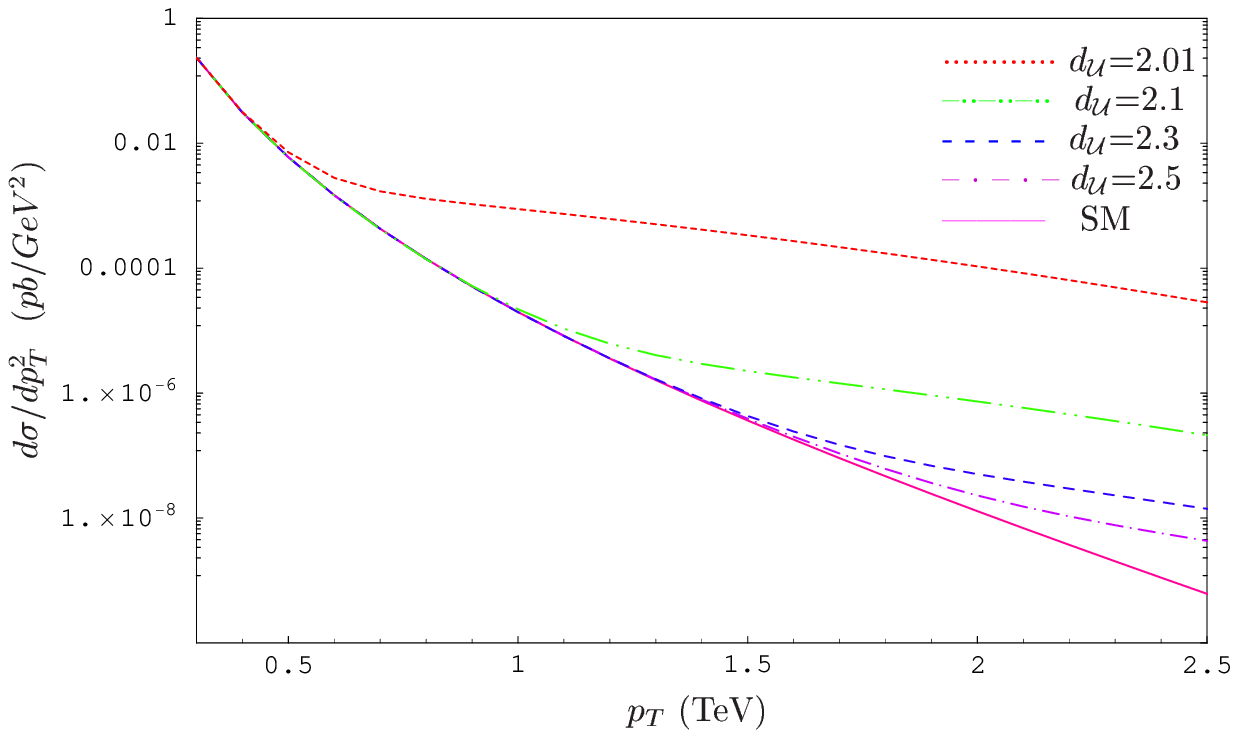}
 \caption{Differential cross section $\frac{d\sigma}{dp_T^2}$ as a function of $p_T$ for two jet signals at the LHC with scalar unparticle
 exchange plus SM contribution. We have set $\lambda_0$=1 and $\Lambda$=1 TeV.}\label{fig2}
\end{figure}
\begin{figure}[htbp!]
 \includegraphics[width=9cm]{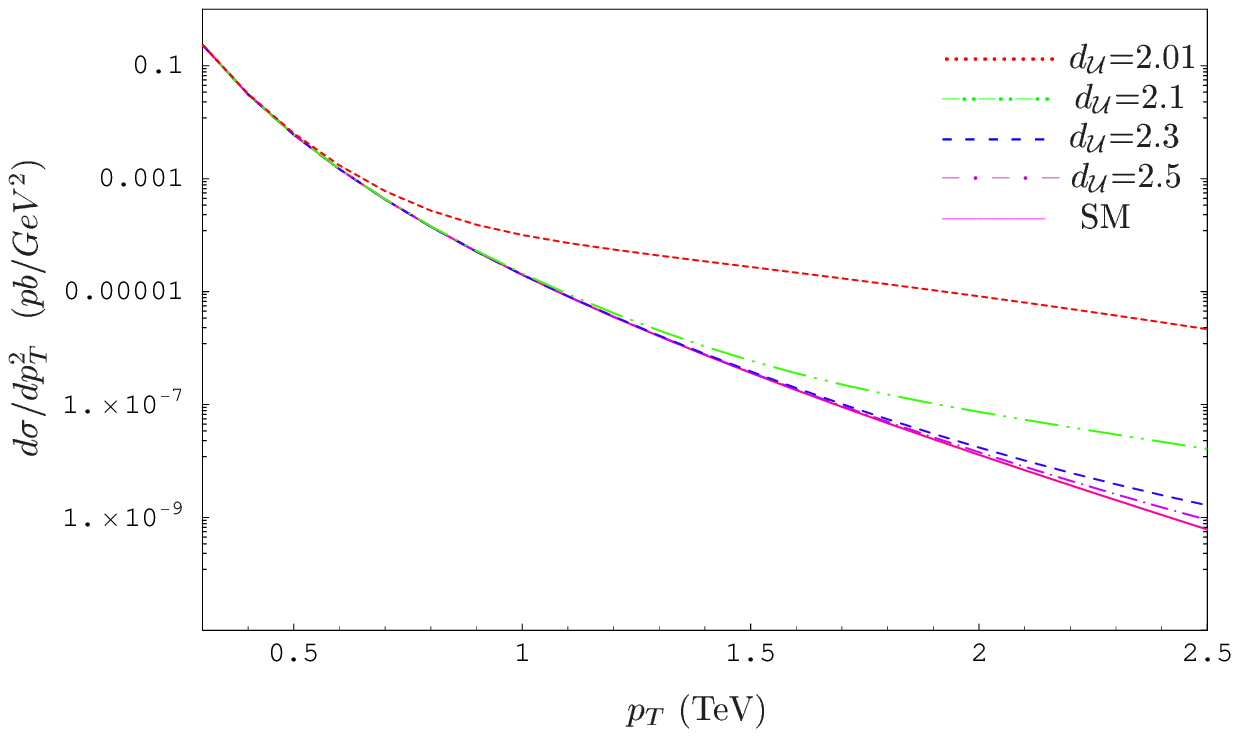}
 \caption{Differential cross section $\frac{d\sigma}{dp_T^2}$ as a function of $p_T$ for two jet signals at the LHC with tensor unparticle
 exchange plus SM contribution. we have set $\lambda_2$=1 and $\Lambda$=1 TeV.}\label{fig4}
\end{figure}
where \[A_T=\frac{\lambda_2^2}{16\lambda_0^2}A_S.\] We have checked
the calculation of the differential cross section for the tensor
unparticle propagator in CFT with $T_{\mu\nu,\rho\sigma}$ given in
Eq.~(\ref{eq6}) and obtained exactly the same result. But in CFT due
to the constrain $d_\U\geq 4$ on the scale dimension for the tensor
unparticle operator, the unparticle exchange contribution to the
hadronic distributions are negligible small. Hence, in the numerical
analysis we have taken the range of $2<d_\U<3$ (conformal invariance
is not imposed further on).

Total hadronic cross section is calculated as a composition of
partonic cross section multiplied by parton densities $f_i(x_i,Q^2)$
which are evaluated at a factorization scale $Q$. The longitudinal
momentum fractions of incoming gluons $x_i=\frac{p_i}{P}$, are
related to the observed jet variables, transverse momenta, $p_{iT}$
and rapitidies, $y_i$ by
\begin{eqnarray}
x_1=\frac{p_T}{\sqrt s}(e^{y_1}+e^{y_2}), x_2=\frac{p_T}{\sqrt
s}(e^{-y_1}+e^{-y_2})
\end{eqnarray}
with the transverse momentum conservation $p_{1T}=p_{2T}=p_{T}$.
Then, the final hadronic state describing the production of
two-gluon jets at LO is specified by the factorization formula

\begin{eqnarray}\label{pt}
\frac{d^3\sigma}{dy_1dy_2dp_T^2}=\sum_{i,j}(\frac{1}{1+\delta_{ij}})\frac{\hat
s}{s}[f_i(x_1,Q^2)f_j(x_2,Q^2)\nonumber\\+f_i(x_2,Q^2)f_j(x_1,Q^2)]\frac{d\hat\sigma}{d\hat
t}\end{eqnarray}

{\it Numerical results and conclusions}. In Figs.~\ref{fig2} and
\ref{fig4} we plot $p_T$ distributions of hadronic cross sections
for two-gluon jets production at the LHC by taking into account the
contributions of the scalar and the tensor unparticles,
respectively, for various $d_{\U}$ values by setting the parameters
$\lambda_0=\lambda_2$=1 and $\Lambda$=1 TeV. We have used CTEQ5
\cite{CTQ5} gluon distributions and have evaluated them at the scale
$Q=p_T$ in all of our numerical computations. Effects of virtual
unparticles become quite large for the $p_T$ values grater than 1
TeV. For instance, for $d_{\U}$=2.01 hadronic differential cross
sections at $p_T$=2 TeV are $10^{-4}$ pb/GeV$^2$ for scalar
unparticle plus SM and $10^{-5}$ pb/GeV$^2$ for the tensor
unparticle plus SM contributions, respectively (the SM value is
$1.2\times10^{-8}$ pb/GeV$^2$). The tensor contribution is about one
order less compared to that of scalar contribution on the average.
These contributions decrease with increasing $d_{\U}$ values. The
corresponding values are $5.10^{-8}$ pb/GeV$^2$ and
$1.7\times10^{-8}$ pb/GeV$^2$, respectively, for $d_\U$=2.3.
\begin{figure}[htbp!]
 \includegraphics[width=9cm]{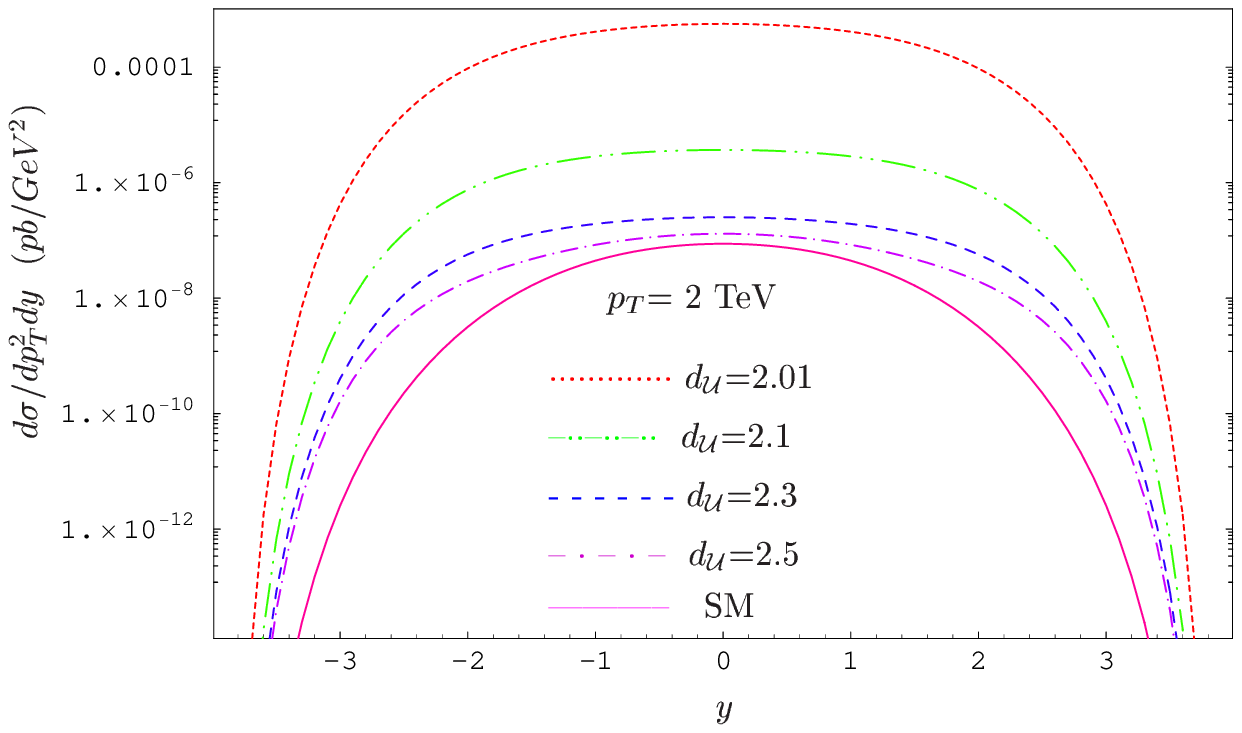}
  \caption{The differential cross section for two gluon jets production at the LHC as a function of pseudo-rapidity interval $y$, at $y_{boost}$=0 and
  $p_T$=2 TeV, for the scalar unparticle exchange plus SM contribution. We have set $\lambda_0$=1 and $\Lambda$=1 TeV}\label{fig3}
\end{figure}
\begin{figure}[htbp!]
 \includegraphics[width=9cm]{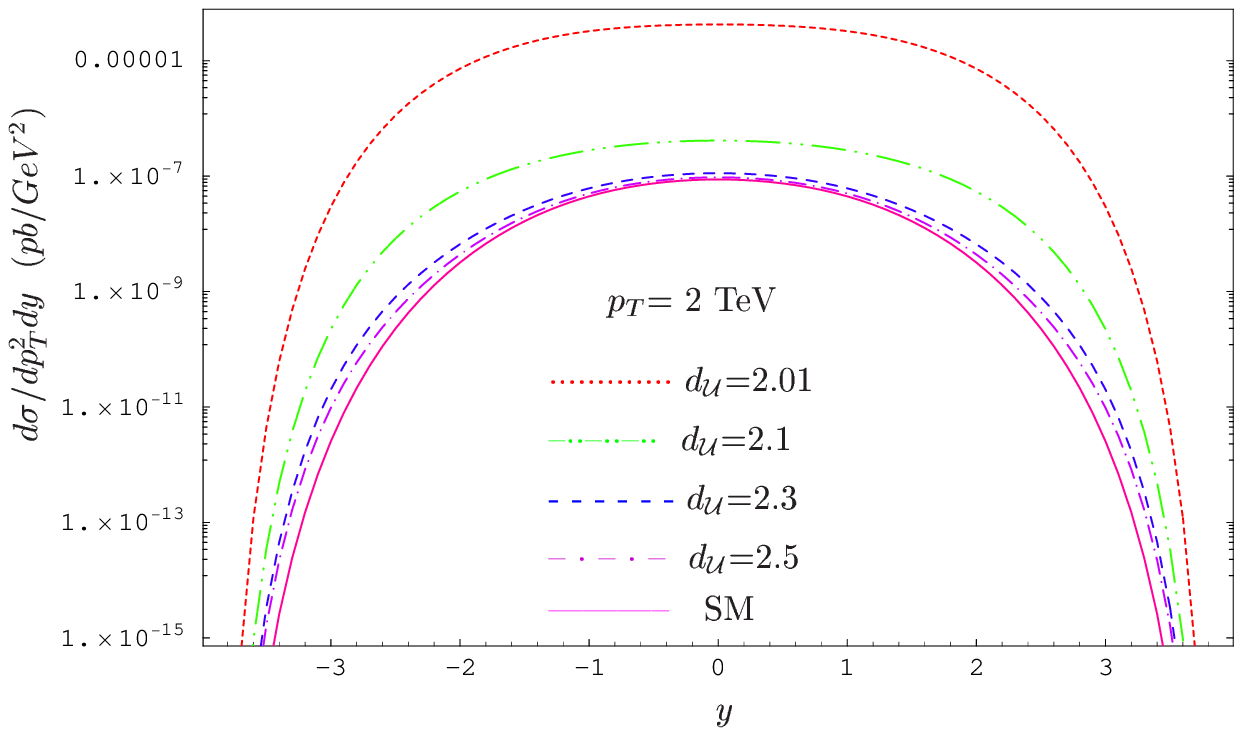}
  \caption{The differential cross section for two gluon jets production at the LHC as a function of pseudo-rapidity interval $y$, at $y_{boost}$=0 and
  $p_T$=2 TeV, for the tensor unparticle exchange plus SM contribution. We have set $\lambda_2$=1 and $\Lambda$=1 TeV}\label{fig5}
\end{figure}
Figs.~\ref{fig3} and \ref{fig5} show the pseudo-rapidity
distributions of cross sections for two-gluon jets production at the
LHC with scalar and tensor unparticle contributions, respectively.
The cross sections are plotted as function of pseudo-rapidity
interval $y=y_1-y_2$ at $y_{boost}=(y_1+y_2)/2$=0. Here, $y$ is the
pseudo-rapidity of a parton in the center of mass frame while
$y_{boost}$ is the pseudo-rapidity of parton center of mass frame
with respect to the hadron center of mass frame. For $d_\U=2.01$ and
the interval $y=1$ the hadronic differential cross sections at
$p_T=2$ are about $4.0\times10^{-4}$ and $3.2\times10^{-5}$
pb/GeV$^2$ for the scalar and the tensor unparticle cases,
respectively. The LO SM differential cross section at the same $y$
and $p_T$ values is $4.5\times10^{-8}$ pb/GeV$^2$. It is possible to
apply an SM $K$ factor to obtain reliable NLO predictions but we
have not included it in our calculations since $K$ values change
depending on the different pseudo-rapidity intervals
\cite{Campbell:2006wx}.

To summarize, we have studied the production of $gg$ jet pairs
associated with unparticle physics at the LHC energies. In the
analysis we have not included $q\bar q$ initial state. The LO SM
contribution originated from $q\bar q$ annihilation at $p_T$=3 TeV
is about 15 \% and it decreases down to about 2 \% at $p_T$=1 TeV
and \% 0.8 at $p_T$=0.5 TeV. For an interval of scale dimension
$2<d_\U<3$ we obtained large enhancements (approximately 10000
times) in $p_T$ and pseudo-rapidity distributions of cross sections;
specifically for $d_{\U}$ values close to 2. It is expected that
even higher enhancement is possible if one extrapolates the interval
to $1<d_\U<2$. We hope that our predictions provide a reliable
explanation for the deviations from SM values of cross section
distributions for $gg$ jet pairs to be produced at the LHC, in the
case of possible existence of unparticle physics.
\begin{acknowledgements}
{\it Acknowledgments.} This work is partially supported by Abant
Izzet Baysal University Research Fund.
\end{acknowledgements}

\end{document}